\documentclass[12pt,preprint]{aastex}
\usepackage{textcomp}
\usepackage{amsmath}
\usepackage{graphicx}
\usepackage[T1]{fontenc}

\def\beq{\begin{eqnarray}}
\def\eeq{\end{eqnarray}}

\begin{document}

\title{Vertical convection in neutrino-dominated accretion flows}

\author{Tong Liu\altaffilmark{1,2,3,4}, Wei-Min Gu\altaffilmark{1,4}, Norita Kawanaka\altaffilmark{5}, and Ang Li\altaffilmark{1,3,4}}

\altaffiltext{1}{Department of Astronomy and Institute of Theoretical Physics and Astrophysics, Xiamen University, Xiamen, Fujian 361005, China; tongliu@xmu.edu.cn}
\altaffiltext{2}{Key Laboratory for the Structure and Evolution of Celestial Objects, Chinese Academy of Sciences, Kunming, Yunnan 650011, China}
\altaffiltext{3}{State Key Laboratory of Theoretical Physics, Institute of Theoretical Physics, Chinese Academy of Sciences, Beijing 100190, China}
\altaffiltext{4}{SHAO-XMU Joint Center for Astrophysics, Xiamen University, Xiamen, Fujian 361005, China}
\altaffiltext{5}{Department of Astronomy, Graduate School of Science, The University of Tokyo, 7-3-1, Hongo, Bunkyo-ku, Tokyo 113-0033, Japan; norita@astron.s.u-tokyo.ac.jp}
\email{tongliu@xmu.edu.cn}

\begin{abstract}
We present the effects of the vertical convection on the structure and luminosity of the neutrino-dominated accretion flow (NDAF) around a stellar-mass black hole in spherical coordinates. We found that the convective energy transfer can suppress the radial advection in the NDAF, and that the density, temperature and opening angle are slightly changed. As a result, the neutrino luminosity and annihilation luminosity are increased, which is conducive to achieve the energy requirement of gamma-ray bursts.
\end{abstract}

\keywords {accretion, accretion disks - black hole physics - convection - gamma-ray burst: general - neutrinos}

\section{Introduction}

The central engine of gamma-ray bursts (GRBs) is usually modelled as the system consisting of a rapidly spinning stellar black hole and an extremely optically thick disk with high density and temperature, namely neutrino-dominated accretion flow (NDAF). The neutrino radiation is the main cooling mechanism, which possibly has the ability to power GRBs via neutrino annihilation outside the disk. A lot of works on this model had been done in the past decades, which can be divided into the following categories: the researches on the stable or time-dependent radial characteristics of the disk, such as the dynamics, structure, components, and luminosity of the NDAF model (see, e.g., Popham et al. 1999; Narayan et al. 2001; Di Matteo et al. 2002; Kohri \& Mineshige 2002; Kohri et al. 2005; Janiuk et al. 2004, 2007; Lee et al. 2005; Gu et al. 2006; Chen \& Beloborodov 2007; Kawanaka \& Mineshige 2007; Liu et al. 2007; Kawabata et al. 2008; Zhang \& Dai 2009; Kawanaka et al. 2013; Li \& Liu 2013; Xue et al. 2013) or magnetic NDAF model (e.g., Lei et al. 2009; Janiuk \& Yuan 2010; Luo et al. 2013), and the corresponding applications to GRBs observations (e.g., Reynoso et al. 2006; Lei et al. 2007; Lazzati et al. 2008; Liu et al. 2010b, 2012b; Romero et al. 2010; Kawanaka \& Kohri 2012; Sun et al. 2012; Hou et al. 2014a, b), on the vertical characteristics of the model, and the applications to observations (e.g., Sawyer 2003; Liu et al. 2008, 2010a, 2012a, 2013, 2014; Pan \& Yuan 2012), and on the time-dependent hydrodynamical simulations of the model and their applications (e.g., Ruffert \& Janka 1999; Lee et al. 2004; Setiawan et al. 2006; Lee et al. 2009; Carballido \& Lee 2011; Sekiguchi \& Shibata 2011; Caballero et al. 2012; Fern{\'a}ndez \& Metzger 2013; Janiuk et al. 2013; Just et al. 2015).

The structures and luminosities of the NDAFs in spherical coordinates had been studied (e.g., Liu et al. 2010a, 2012a, 2013). We stressed that the geometrical thickness of the disk was relevant to the advection and showed the existence of an empty funnel along the rotation axis which makes neutrino annihilation more efficient (Liu et al. 2010a). Furthermore, the distribution of the Bernoulli parameter, neutrino trapping, and components features, i.e., nucleosynthesis near the surface of the disk, were studied in the vertical direction (Liu et al. 2012a, 2013). The possible outflow might occur at the outer region of the disk, and the heavy nuclei tended to be produced near the surface of the disk, especially for the outer region. We also argued that the self-gravity effect may be important to the structure of the NDAF, and the neutrino luminosity further (Liu et al. 2014). The gravitational instability of the disk was reviewed, which might account for the X-ray flares or extended emission in GRBs.

Convective motions have been widely studied in the black hole accretion disks. The vertically convective motions had been considered especially for the radiation pressure-dominated accretion disks (see, e.g., Bisnovatyi-Kogan \& Blinnikov 1977; Shakura et al. 1978; Milsom et al. 1994; Blaes \& Socrates 2001; S{\c a}dowski et al. 2009, 2011; Gu 2012, 2015; Kawanaka \& Kohri 2012). If the convective energy transfer in the vertical direction becomes effective, the inwardly advective energy would be suppressed. The convective motion in an accretion disk becomes important when the radiation pressure is dominant. As we know, the typical mass accretion rate of NDAF is about $0.1-1$ $M_\odot ~\rm s^{-1}$, and the outer region of the NDAF should be radiation pressure-dominated if the convection is not taken into account (e.g., Liu et al. 2007; Kawanaka \& Mineshige 2007). In the innermost region, the neutrino cooling becomes dominant, and the pressure would be dominated by gas. Thus we can study the convection motions in the NDAF directly. Kawanaka \& Kohri (2012) first investigated the effects of the vertical convection in the NDAF model. If the vertical convection is considered, the hyperaccretion disk is expected to be hotter and the neutrino emission due to the electron-positron annihilation becomes the most efficient cooling process. They also presented that the sequence of the thermal equilibrium solutions for the convective NDAF had a viscously unstable branch, especially for the relatively small viscosity parameter, which might explain the origin of the drastic variabilities in the prompt emissions of GRBs. Recently, Jiang et al. (2014) investigated the super-Eddington accretion flows onto black holes using a global three-dimensional radiation magneto-hydrodynamical simulation. They found that the radiation of slim disk (e.g., Abramowicz et al. 1988) can be greatly increased via the vertical advection of radiation caused by magnetic buoyancy. As the natural extension of the slim disk in the state for very high accretion rate (Liu et al. 2008), NDAF should be expected to radiate a larger number of neutrinos due to the same mechanism.

In this paper, we introduce the vertical convection in the thermodynamics of the NDAF model, and present its effects on the structure and luminosity of the hyperaccretion disk in spherical coordinates. The basic equations, boundary condition, and convective effect are presented in Section 2. In Section 3, the corresponding results of the structure, neutrino luminosity and annihilation luminosity, and comparisons with the cases excluding the convection are shown. Summary is made in Section 4.

\section{Equations}
\subsection{Basic equations}

As shown in Figure 1, a steady state axisymmetric accretion flow in spherical coordinates ($R$, $\theta$, $\phi$), i.e., $\partial/\partial t =\partial/\partial \phi = 0$ is considered (e.g., Kato et al. 2008). We adopt the Newtonian potential since it is convenient for self-similar assumptions. The basic equations of continuity and momentum are the following (see, e.g., Xue \& Wang 2005; Gu et al. 2009):
\beq \frac{1}{R^2} \frac{\partial}{\partial R}(R^2 \rho v_R) +\frac{1}{R^2 \sin \theta} \frac{\partial}{\partial \theta} (\sin \theta \rho v_\theta) = 0, \eeq
\beq v_R \frac{\partial v_R}{\partial R} + \frac{v_\theta}{R} (\frac{\partial v_R}{\partial \theta} - v_\theta) - \frac{{v_\phi }^2}{R} = -\frac{GM}{R^2} -\frac{1}{\rho} \frac{\partial p}{\partial R}, \eeq
\beq v_R \frac{\partial v_\theta}{\partial R} + \frac{v_\theta}{R}(\frac{\partial v_\theta}{\partial \theta} + v_R) - \frac{{v_\phi}^2}{R} \cot \theta = -\frac{1}{\rho R} \frac{\partial p}{\partial\theta}, \eeq
\beq v_R \frac{\partial v_\phi}{\partial R}+\frac{v_\theta}{R} \frac{\partial v_\phi}{\partial \theta}+\frac{v_\phi}{R} (v_R + v_\theta \cot \theta) = \frac{1}{\rho R^3}\frac{\partial}{\partial R} (R^3 T_{R \phi}), \eeq
where $v_R$, $v_{\theta}$, and $v_{\phi}$ are the three components of the velocity, and $p$ and $\rho$ are the pressure and density of the disk. Here, the $R\phi$-component of the viscous stress tensor are considered, i.e., $T_{R\phi} = \rho \nu R \partial (v_{\phi}/R)/\partial R$. The kinematic coefficient of viscosity is written as $\nu = \alpha c_s^2  / \Omega_{\rm K}$, where the sound speed $c_s$ is defined as $c_s^2 = p/\rho$, the Keplerian angular velocity is $\Omega_{\rm K} = (GM/R^3)^{1/2}$, $M$ is the mass of the black hole, $v_\phi= R \Omega_{\rm K}$, and $\alpha$ is a constant viscosity parameter. Following Narayan \& Yi (1995), the radial self-similar assumption is
\beq \rho(R, \theta) \propto R^{-3/2} \rho (\theta), \eeq
\beq c_s (R, \theta) \propto R^{-1/2} c_s (\theta), \eeq
\beq v_R (R, \theta) \propto R^{-1/2}v_R (\theta), \eeq
\beq v_\phi (R, \theta) \propto R^{-1/2} v_\phi (\theta),\eeq
\beq v_\theta (R, \theta)= 0. \eeq

According to substitute the above equations into Equation (1)-(4), which are transformed to the ordinary differential equations with respect of $\theta$. Thus the continuity equation, i.e., Equation (1), is an identical equation, but we can define the mass accretion rate $\dot{M}$ to replace it as
\beq \dot{M} = -4 \pi R^2 \int_{\theta_0}^{\pi /2} \rho v_R \sin \theta d \theta, \eeq
where $\theta_0$ is the polar angle of the surface as shown in Figure 1, and the other physical quantities with subscripts ``0'' represent the values at the equatorial plane of the disk.

Furthermore, the momentum equations, i.e., Equation (2)-(4), can be simplified as (e.g., Gu et al. 2009; Liu et al. 2010a, 2012a)
\beq \frac{1}{2} {v_R}^2 + \frac{5}{2} {c_{\rm s}}^2 + {v_\phi}^2 - R^2 {\Omega_{\rm K}}^2 = 0, \eeq
\beq \frac{1}{\rho} \frac{d p}{d \theta} = R^2 {\Omega_{\rm K}}^2 \cot \theta, \eeq
\beq v_R = -\frac{3}{2} \frac{\alpha {c_{\rm s}}^2}{R \Omega_{\rm K}}. \eeq

The equation of state is
\beq p &=& p_{\rm gas} + p_{\rm rad} + p_{\rm e} + p_\nu = \frac{\rho k T}{m_{\rm p}} (\frac{1+3X_{\rm nuc}}{4}) + \frac{11}{12} aT^4 + \frac{2\pi h c}{3} (\frac{3 Y_{\rm e} \rho}{8\pi m_{\rm p}}) + \frac{\mu_\nu}{3}, \eeq
where $p_{\rm gas}$, $p_{\rm rad}$, $p_{\rm e}$, and $p_\nu$ are the gas pressure from nucleons, the radiation pressure of photons, the degeneracy pressure of electrons, and the radiation pressure of neutrinos, respectively. $k$, $T$, $m_{\rm p}$, $a$, and $h$ are the Boltzmann constant, temperature of the disk, proton mass, radiation constant and Planck constant, respectively. The equation can be applied to anywhere on the disk to describe the pressure, but we also need an equation to completely give the vertical distributions of the density, temperature, and pressure. For simplicity, we assume the polytropic relation, i.e., $p=K\rho^{4/3}$, in the vertical direction, $K$ is a constant (e.g., Lee et al. 2005; Liu et al. 2008, 2010a, 2012a, 2013). Here we also assume the electron fraction $Y_{\rm e}$ is 0.5 (e.g., Di Matteo et al. 2002; Gu et al. 2006; Liu et al. 2014), because its influences are very limited on the structure and neutrino luminosity of the disk, unless we care about the flavors of neutrinos emitted from the disk (e.g., Liu et al. 2013). $X_{\rm nuc}$ is the mass fraction of free nucleons, and its expression obtained by taking $Y_{\rm e} = 0.5$ is (e.g., Kohri et al. 2005)
\beq X_{\rm nuc}= \min [1,~295.5 \rho_{10}^{-3/4} T_{11}^{9/8} \exp (-0.8209/T_{11})], \eeq
where $\rho_{10}= \rho /10^{10} ~\rm g~cm^{-3}$ and $T_{11} = T / 10^{11} ~\rm K$. $\mu_\nu$ is the energy density of neutrinos, which can be expressed as (e.g., Lee et al. 2005; Liu et al. 2012a)
\beq  \mu_{\rm \nu}\simeq \sum_{i} \frac{21}{8} a T^4 [1-\exp(-\tau_{{\nu}_i})], \eeq
where $\tau_{{\nu}_i}=\tau_{a,{\nu}_i}+ \tau_{s,{\nu}_i}$ is the total optical depth for neutrinos, $\tau_{a,{\nu}_i}$ and $\tau_{s,{\nu}_i}$ are the absorption and scattering optical depth for neutrinos, and the subscript $i$ runs for the three species of neutrinos $\nu_{\rm e}$, $\nu_{\rm \mu}$, and $\nu_{\rm \tau}$, which can be defined as (Di Matteo et al. 2002; Lee et al. 2005; Liu et al. 2012a)
\beq \tau_{a, \nu_i} \approx \int_{\theta_0}^{\theta}\frac{ 2 q_{\nu_i} R d \theta}{7 \sigma T^4}, \eeq
\beq \tau_{s, \nu_i} \approx 2.7 \times 10^{-7} \int_{\theta_0}^{\theta} \rho_{10} T _{11}^2 R d \theta, \eeq
where $q_{\nu_i}$ is the summation of the cooling rates per unit volume due to the neutrino reactions for different neutrinos, especially the Urca processes and electron-positron pair annihilation (e.g., Di Matteo et al. 2002; Liu et al. 2007). The neutrino scattering processes includes scattering by free protons, free neutrons, $\alpha$-particles, and electrons (e.g., Di Matteo et al. 2002).

The energy equation is written as
\beq Q_{\rm vis} = Q_{\rm adv} + Q_\nu, \eeq
where $Q_{\rm vis}$, $Q_{\rm adv}$, and $Q_\nu$ are the viscous heating, advection and neutrino cooling rates per unit area, respectively (e.g., Liu et al. 2010a, 2012a, 2013). The heating due to the nuclear reaction can be neglected, because the gravitational energy is more powerful and efficient in the accretion systems, even for the hot dense matters in the NDAF.

$Q_{\rm vis}$ and $Q_{\rm adv}$ can be expressed as
\beq Q_{\rm vis} = 2 R \int_{\theta_0}^{\pi/2} q_{\rm vis} \sin {\theta} d\theta, \eeq
\beq Q_{\rm adv} = 2 R \int_{\theta_0}^{\pi/2} q_{\rm adv} \sin {\theta} d\theta, \eeq
where $q_{\rm vis}$ and $q_{\rm adv}$ are the viscous heating and advective cooling rates per unit volume. The viscous heating and advective cooling rate per unit volume are expressed as $q_{\rm vis} = \nu \rho R^2 [\partial (v_{\phi}/R)/\partial R]^2$ and $q_{\rm adv} = \rho v_R [\partial e/\partial R - (p/\rho^2) \partial \rho/\partial R]$ ($e$ is the internal energy per unit mass), respectively. After self-similar simplification, the expression is (Liu et al. 2010a, 2012a)
\beq q_{\rm vis} = \frac{9}{4} \frac{\alpha p v_{\phi}^2}{R^2 \Omega_{\rm K}}, \eeq
\beq q_{\rm adv} = - \frac{3}{2} \frac{(p-p_{\rm e}) v_R}{R}, \eeq
here the entropy of degenerate particles is neglected.

The neutrino radiation cooling rate $Q_\nu$ can be defined as (Lee et al. 2005; Liu et al. 2012a, 2013)
\beq Q_\nu = 2 R \sum_i \int_{\theta_0}^{\pi/2} q_{\nu_i} {\rm e}^{-\tau_{\nu_i}} \sin {\theta} d\theta, \eeq
As mentioned, different types of neutrinos are involved into different types of reactions and have their own optical depths. Additionally, we should point out that the prescription of the energy equation with convection is enough to describe the vertical energy transport (Liu et al. 2010a; Pan \& Yuan 2012).

Additionally, we follow the solutions of the boundary condition in Liu et al. (2012a, 2013). As well as the principle of the Eddington luminosity, based on the balance between the vertical component of the gravity and radiation pressure at the surface of the disk, the surface temperature $T_{\rm s}$ can be defined as
\beq T_{\rm s}=(\frac{6 G M m_{u}}{a \sigma_{\rm T} R^2} \rm cot \theta_0)^{\frac{1}{4}}, \eeq
where $m_{\rm u}$ is the mean mass of a nucleon, $\sigma_{\rm T}$ is the Thompson scattering cross, and $a$ is the radiation density constant. This equation is used to determine the angle of the disk surface and complete the basic equations, i.e., Equations (10)-(14) and (19).

\subsection{Vertical Convection}

In the vertical direction of the disk, the convective motion can carry the photons (or neutrinos) from the equatorial plane to the surface of the disk by the magnetic buoyancy. For slim disk model, the key issue is whether the vertical radiation transports process due to the magnetic buoyancy is faster than the radial advection process, thus the photons can escape from the surface before being advected into the black hole. In Jiang et al. (2014), the results indicated that the radiation can be widely increased because of the vertical advection of radiation caused by magnetic buoyancy. The average advective energy transport velocities along the vertical and radial directions at each radius had been calculated, which are with nonzero values and analogous to the bulk velocities, such as $v_R$, $c_{\rm s}$ or $v_\phi$ (e.g., S{\c a}dowski et al. 2011; Jiang et al. 2014). As mentioned above, should NDAF be expected to radiate a larger numbers of neutrinos due to the same mechanism?

Following the discussion in Kawanaka \& Kohri (2012), we adopt the maximal value of the convective velocity physically allowed. Then we also estimate the typical timescale for the convective motion of the blob in the vertical direction $t_{\rm c}$ and the advection timescale $t_{\rm adv}$. The vertical convective speed along the vertical direction can be approximatively written as
\beq v_{\rm c} \simeq  - v_R \cos\theta. \eeq
Here the components of the bulk velocity $v_R$ are cautiously chosen to present the radial advection and vertical convection (or advection), because there is no references to describe vertical convection well and we further consider the dynamical requirement. This is the vertically ``infalling'' speed onto the equatorial plane due to the vertical component of the gravity of the black hole. In other words, the convective velocity should be the same order of magnitude as the vertical components of $v_R$, otherwise, the convection may be destroyed by the gravity and cannot exist in the vertical direction. Thus the convective timescale is
\beq t_{\rm c} \simeq \int_{\theta_0}^{\theta} \frac{R_0 d \cot \theta'}{v_{\rm c}} \simeq  \int_{\theta_0}^{\theta} \frac{R_0 d \theta'}{v_R \sin^2 \theta' \cos \theta'}, \eeq
where $R_0$ is the radius at the equatorial plane.

Meanwhile, the advection timescale can be expressed as
\beq t_{\rm adv} \simeq - \int_{3R_g}^{R} \frac{d {R}'}{v_R} - \frac{3R_g}{v_R|_{R=3R_g}}, \eeq
where $R_g=2GM/c^2$ is the Schwarzschild radius.

The vertical convection should exist if $t_{\rm c}<t_{\rm adv}$, i.e., the polar angle of the convective region should satisfy $\theta_0<\theta<\theta_{\rm c}$, where the critical polar angle $\theta_{\rm c}$ is defined by
\beq R \sin \theta_{\rm c} \int_{\theta_{\rm c}}^{\theta_0} \frac{d \theta}{v_R \sin^2 \theta \cos \theta} = \int_{3R_g}^{R} \frac{d R'}{v_R} + \frac{3R_g}{v_R|_{R=3R_g}}. \eeq

When the vertical convection is included in the NDAF model and $\theta_0 < \theta_{\rm c}$, Equation (21) can be replaced by
\beq Q_{\rm adv} = 2 R \int_{\theta_{\rm c}}^{\pi/2} q_{\rm adv} \sin {\theta} d\theta, \eeq
otherwise, the equation remains unchanged. Although only one equation is replaced, the essentials of the question has changed, i.e., how much ability of the vertical convection could suppress the radial advection and increase the vertical radiation transports?

\section{Numerical results}

The system of equations is closed, as Equations (10)-(14) and (19) with the boundary condition and given constant parameters $R$, $M$, $\dot{M}$, and $\alpha$. The main difference between the cases including and excluding the vertical convection is the description of the advection cooling rate. In our calculations we take $M = 3 M_\odot$, and define the dimensionless radius and mass accretion rate as $r=R/R_g$ and $\dot{m}=\dot{M}/M_\odot~\rm s^{-1}$, respectively.

\subsection{Structure}

Figure 2 (a) and (b) show the variations of the density and temperature at the equatorial plane with the dimensionless radius $r$, for which the given parameters are $\dot{m}=1$ and $\alpha=0.1$. The solid and dashed lines correspond to the cases including and excluding the vertical convection, respectively. We notice that the variations of the density and temperature in the case excluding the vertical convection are similar to the solutions in Liu et al. (2010a, 2012a, 2013). Considering the convection, the density slightly decreases and the temperature conversely increases, which are consistent with the results in Kawanaka \& Kohri (2012). If the hyperaccretion matter becomes hotter, the thickness of the disk is expected to be geometrically thicker due to the the vertical polytropic distribution of the temperature.

Figure 3 shows the variations of the half-opening angle $(\pi /2 -\theta)$ with the dimensionless radius $r$ for the different cases. The black, red, and blue lines describe the cases that the given parameters ($\dot{m}$, $\alpha$) is (1, 0.1), (1, 0.05), and (0.1, 0.1), respectively. The solid and dashed lines correspond to $(\pi /2 -\theta_0)$ and $(\pi /2 -\theta_{\rm c})$, and the convective motions exist in the area between these two lines to replace the radial advection. The thick and thin black solid lines correspond to the half-opening angle of the disk in the cases including and excluding the vertical convection. We also notice that the half-opening angle of the disk in the case excluding the vertical convection are similar to the solutions in Liu et al. (2012a, 2013), which has the slight difference comparing with the case including the convection.

As the solutions in Liu et al. (2007, 2010a), the advection and neutrino cooling mechanisms are respectively dominant cooling modes in the outer and inner region of the disk. Thus the effects of the vertical convection mainly reflect in the inner region, i.e., $\theta_{\rm c}$ is closed to $\pi/2$, until very near the black hole, $R \leq 10R_g$. The amplitudes of the differences of the density and temperature between the cases including and excluding the convection can also be understood by the above explanations.

Besides, we also notice that the accretion rate and viscous parameter can markedly change the polar angle $\theta_0$ and $\theta_{\rm c}$, and the effects of the viscous parameter are much more significant. Higher density and temperature is conducive to the more neutrinos emission, and the density and temperature decrease gradually from the equatorial plane to the surface of the disk, so the number of neutrinos radiated from the equatorial plane of the disk are much more than that from other regions, especially nearby the surface. The convective energy transfer in the vertical direction can be effective to suppress the radial advection in the NDAF, but the region dominated by the vertical convection is deviated from the equatorial plane of the disk, thus it can be expected that the neutrino emission rate should increase slightly.

\subsection{Neutrino Luminosity and annihilation luminosity}

Using the neutrino cooling rate $Q_\nu$, the neutrino luminosity before annihilation $L_\nu$ is obtained as
\beq L_ \nu = 4 \pi \int_{3R_g}^{300R_g} Q_\nu R d R. \eeq
Furthermore, in Liu et al. (2010a), the luminosity of neutrino annihilation $L_{\nu \bar{\nu}}$ can be roughly evaluated by the assumption: $\eta \propto V_{\rm ann}^{-1}$ (see, e.g., Mochkovitch et al. 1993), where $\eta \equiv L_{\nu \bar{\nu}}/L_\nu$ is the annihilation efficiency, and $V_{\rm ann}$ is the volume above the disk. If we take the outer boundary of the disk $R_{\rm out}$ is $300R_g$, $V_{\rm ann}$ can be estimated by integrating the region of $\theta < \theta_0$ and $R < R_{\rm out}$. We can estimate the proportionality constant of the efficiency to the volume above the disk as a function of accretion rate from the vertically-integrated NDAF model where $\theta_0$ corresponds to $\pi/2$ (e.g., Liu et al. 2007).

Figure 4 displays the neutrino luminosity $L_\nu$ and annihilation luminosity $L_{\nu \bar{\nu}}$ for the varying dimensionless mass accretion rates $\dot{m}$. The solid and dashed lines correspond to the cases including and excluding the vertical convection, respectively. The thick and thin lines display the estimations of the neutrino luminosity and annihilation luminosity, respectively. The luminosity has a certain increase, which is origin from the suppressed advection and correspondingly increased neutrino cooling in all radii of the disk. The influence on the inner region of the disk is stronger in the cases of the lower accretion rate (e.g., $\dot{m} =0.1$) than in the cases of larger accretion rate (e.g., $\dot{m} =1$) as shown in Figure 3. Because neutrinos are mainly launched from the inner region of the disk, the difference of the luminosity between with and without the convection in the case of the lower accretion rate is slightly larger than that in the case of the larger accretion rate. Additionally, in Kawanaka \& Kohri (2012), they found that the electron-positron pair annihilation replaces the Urca process as a dominant neutrino emission process when convection works efficiently. These two processes are intensely related to the density and temperature (e.g., Di Matteo et al. 2002; Liu et al. 2007). In our calculations, because the density and temperature do not change apparently, the annihilation becomes stronger but still weaker than the Urca processes until for the larger radius.

Considering with the increase of the opening angle, the neutrino annihilation luminosity will also be magnified as shown in the Figure 4. Obviously, the neutrino luminosity and annihilation luminosity derived from the calculations considering the effects of the opening angle of the disk are significantly larger than those derived from the calculations neglecting the thickness of a disk as shown in the above estimation method of annihilation luminosity (e.g., Liu et al. 2010). We notice that for $\dot{M}\sim 5~M_\odot~{\rm s}^{-1}$, the density of radiated neutrino is so large that the annihilation efficiency is close to 1. Furthermore, the disk is extremely thick, which restrains the ejection existing in a vary narrow empty funnel along the rotation axis. As discussed in Liu et al. (2010a), since the large opening angle of the disk and the sufficient output of energy, the neutrino annihilable ejection required by GRBs may be naturally explained.

\section{Summary}

For one of the radiation pressure-dominated accretion disks, we study the effects of the vertical convection on the structure and luminosity of the NDAF around a stellar-mass black hole in spherical coordinates. The major points we wish to stress are as follows:

\begin{enumerate}
\item
The convective energy transfer in the vertical direction can be effective to suppress the radial advection for the different accretion rates and viscous parameters in all radii of the NDAF, but the region dominated by the vertical convection is deviated from the equatorial plane of the disk.

\item
The density, temperature, and opening angle of the disk are slightly different between two cases with and without the vertical convection.

\item
Considering the vertical convection, the neutrino luminosity and annihilation luminosity have a certain increase. The increased opening angle of the disk can enhance the annihilable efficiency and the collimation of the ejecta, which further verifies that the NDAF can be regarded as the candidate of the central engine of GRBs.
\end{enumerate}

In future, we should work on the numerical calculations or simulations of the NDAF model with more precise description of the vertical convection and the neutrino transfer to replace the rough estimates by the convective region and polytropic relation in this work.

\begin{acknowledgements}
This work was supported by the National Basic Research Program of China (973 Program) under grant 2014CB845800, the National Natural Science Foundation of China under grants 11222328, 11233006, 11333004, 11373002, 11473022, U1331101, and U1431107, and the CAS Open Research Program of Key Laboratory for the Structure and Evolution of Celestial Objects under grant OP201305.
\end{acknowledgements}

\clearpage

\begin{figure}
\centering
\includegraphics[width=0.8\textwidth]{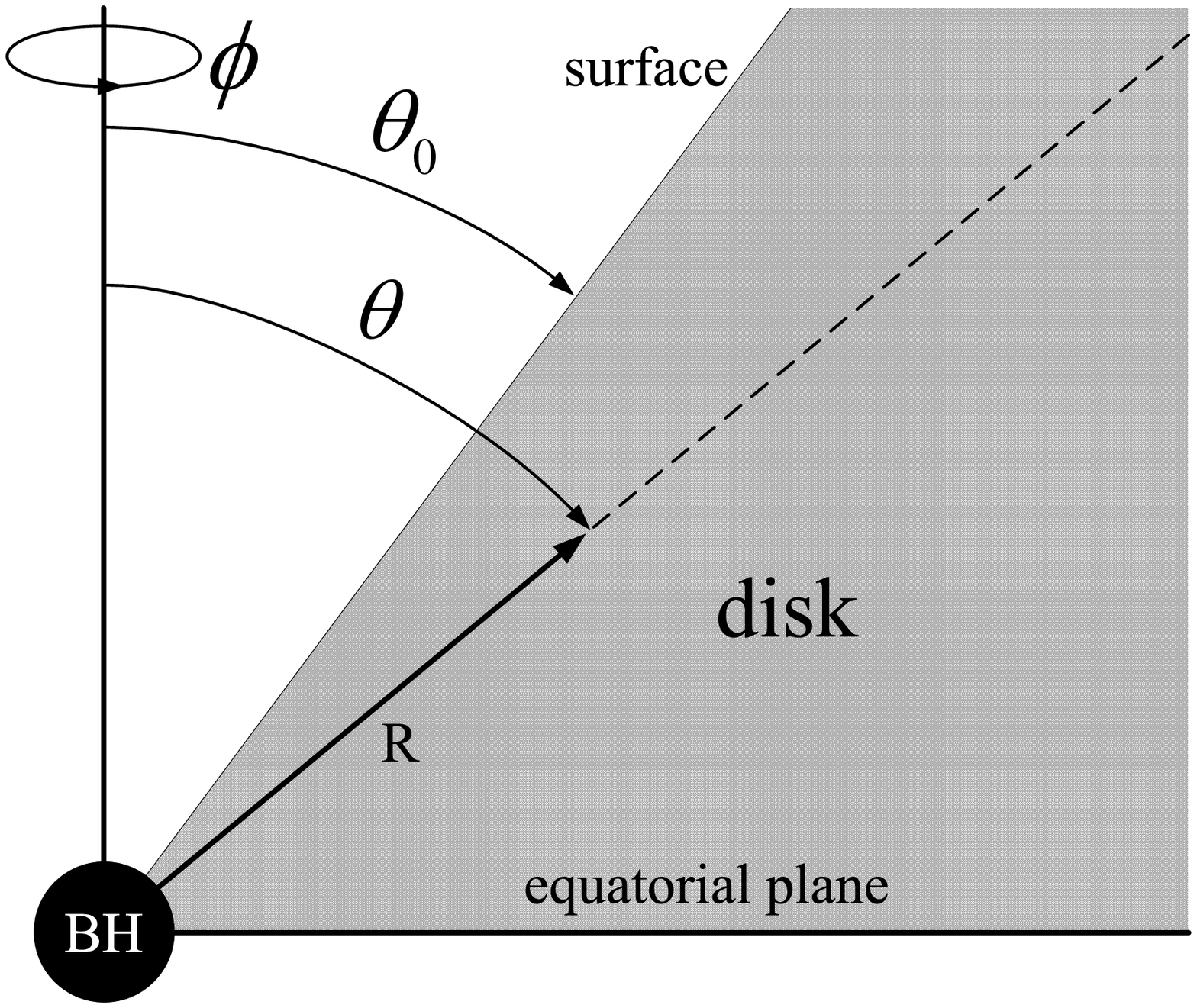}
\caption{Schematic picture of the system composed of a black hole and an accretion disk in spherical coordinates, similarly see also in Liu \& Xue (2011).}
\label{sample-figure1}
\end{figure}

\clearpage

\begin{figure}
\centering
\includegraphics[angle=0,scale=0.45]{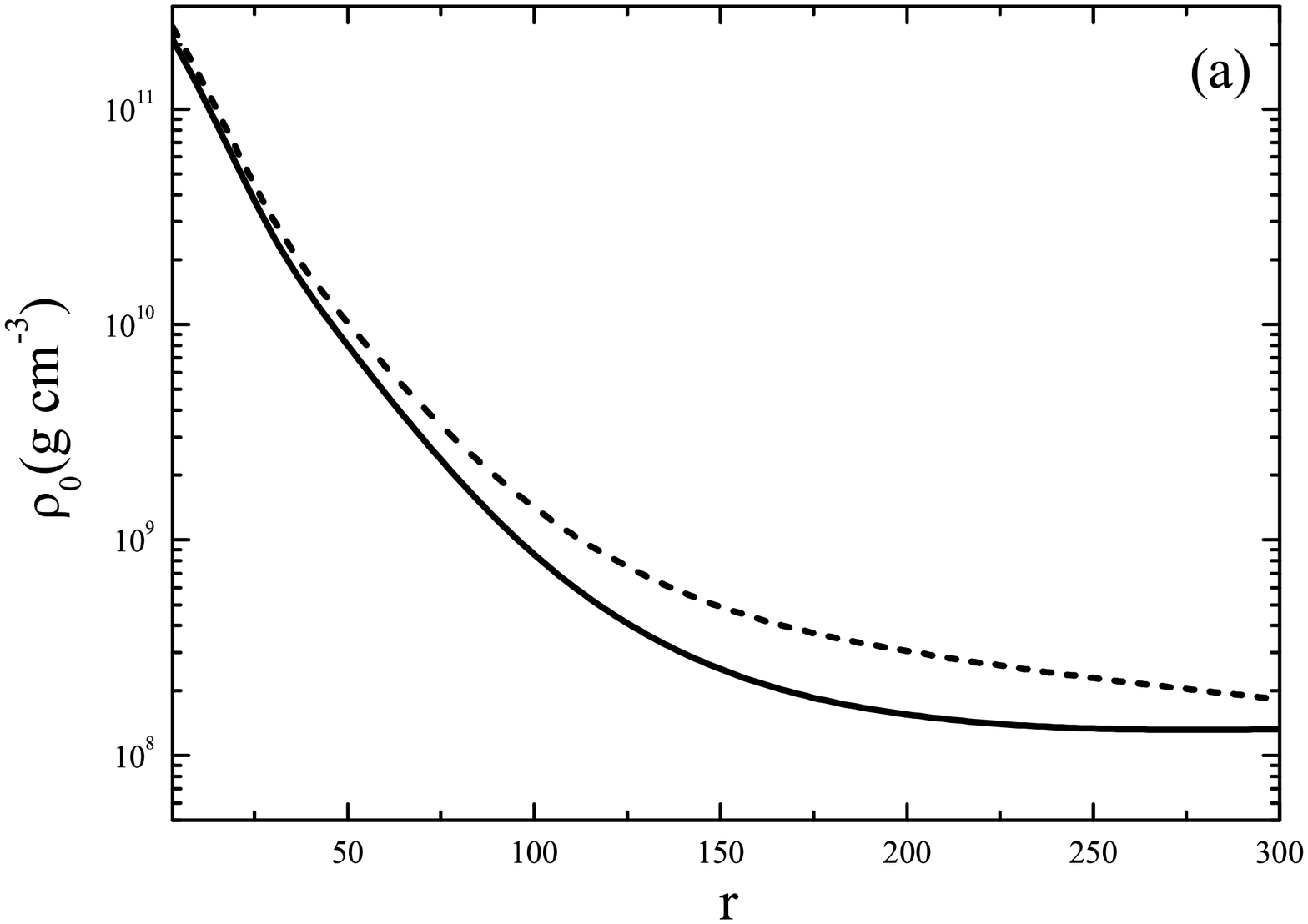}
\includegraphics[angle=0,scale=0.45]{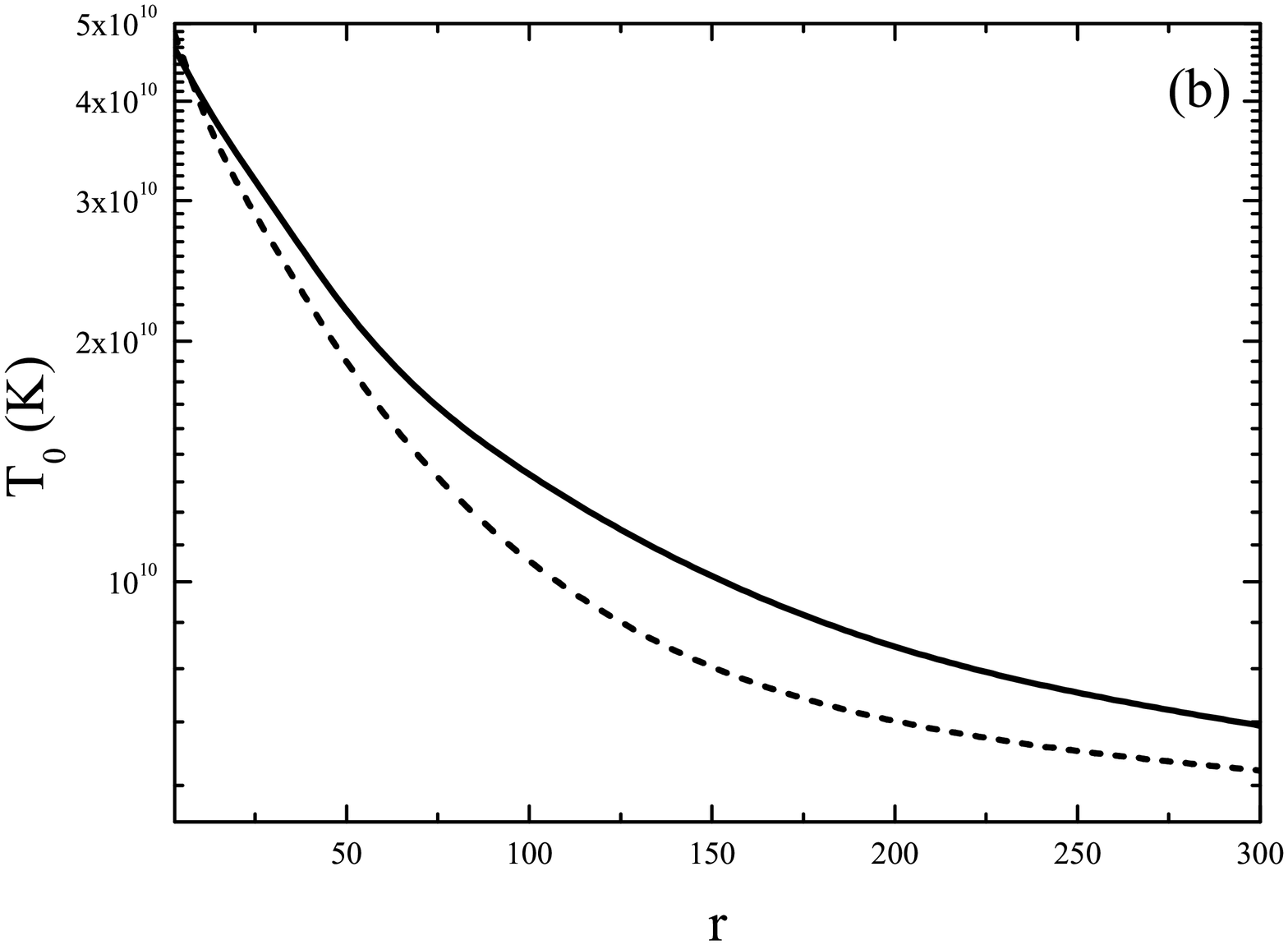}
\caption{Variations of the density and temperature at the equatorial plane ($\rho_0$ and $T_0$) with the dimensionless radius $r$, for which the given parameters are $\dot{m}=1$ and $\alpha=0.1$. The solid and dashed lines correspond to the cases including and excluding the vertical convection, respectively.}
\label{sample-figure2}
\end{figure}

\clearpage

\begin{figure}
\centering
\includegraphics[angle=0,scale=0.6]{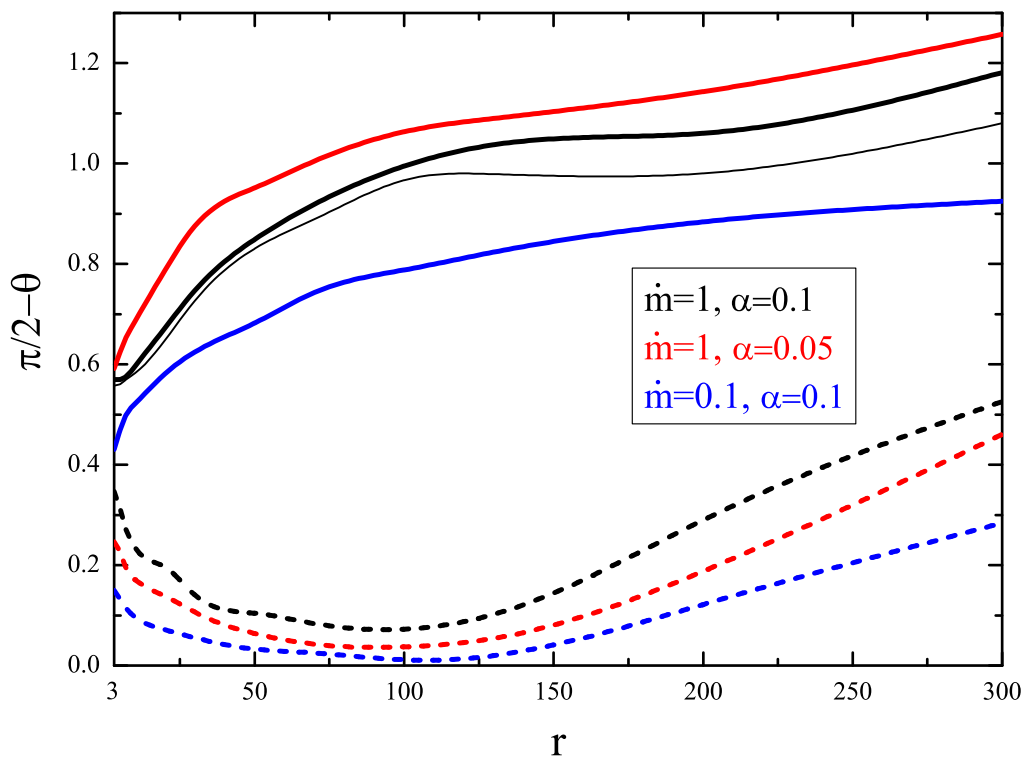}
\caption{Variations of the half-opening angle $(\pi /2 -\theta)$ with the dimensionless radius $r$ for the different cases. The solid and dashed lines correspond to $(\pi /2 -\theta_0)$ and $(\pi /2 -\theta_{\rm c})$, respectively. The thick and thin black solid lines correspond to the half-opening angle of the disk in the cases including and excluding the convection with $\dot{m}=1$ and $\alpha=0.1$.}
\label{sample-figure3}
\end{figure}

\clearpage

\begin{figure}
\centering
\includegraphics[angle=0,scale=0.6]{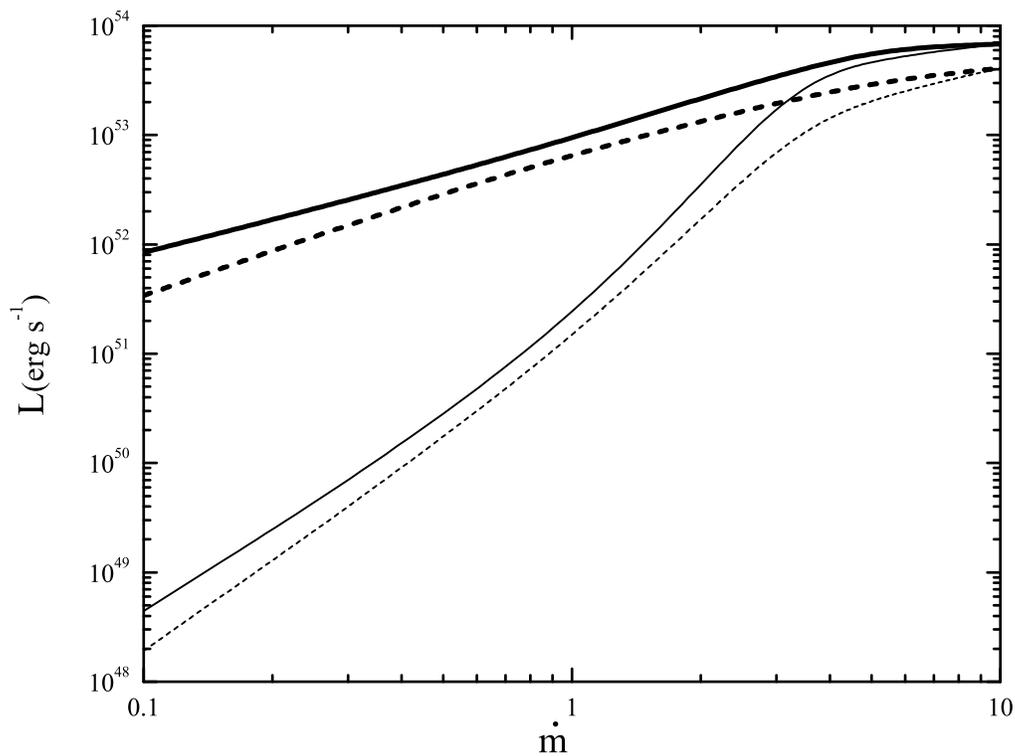}
\caption{Neutrino luminosity $L_\nu$ and annihilation luminosity $L_{\nu \bar{\nu}}$ for varying dimensionless mass accretion rates $\dot{m}$. The solid and dashed lines correspond to the cases including and excluding the vertical convection, respectively. The thick and thin lines display the estimations of the neutrino luminosity and annihilation luminosity, respectively.}
\label{sample-figure4}
\end{figure}

\clearpage

\end{document}